\documentclass{article}
\usepackage[T1]{fontenc}
\usepackage[latin1]{inputenc}
\usepackage{geometry}
\usepackage{graphics}
\usepackage{setspace}
\usepackage{subfigure}

\makeatletter

\providecommand{\LyX}{L\kern-.1667em\lower.25em\hbox{Y}\kern-.125emX\@}

 \newcommand{\lyxaddress}[1]{
   \par {\raggedright #1 
   \vspace{1.4em}
   \noindent\par}
 }

\makeatother

\begin{document}

\title{{\huge Dynamical Scaling In Two Dimensional Quenched Uniaxial Nematic Liquid
Crystals}\huge }

\date{~~~~~~~~~}

\author{Subhrajit Dutta\thanks{
email: subhro@juphys.ernet.in
} and Soumen Kumar Roy\thanks{
email : skroy@juphys.ernet.in
}}

\maketitle

\lyxaddress{~~~~~~~~~~~~~~~~~~~~~~~~~~~~~~Department of Physics,
Jadavpur University, Calcutta-700 032, India.}

\begin{abstract}
{\large The phase ordering kinetics of the two-dimensional uniaxial nematic
has been studied using a Cell Dynamic Scheme. The system after quench from T=\( \infty  \)
was found to scale dynamically with an asymptotic growth law similar to that
of two-dimensional O(2) model (quenched from above the Kosterlitz - Thouless
transition temperature), i.e. L(t)\( \sim  \)(t/ln(t/t\( _{0} \)))\( ^{1/2} \)(with
nonuniversal time scale t}\textbf{\large \( _{0} \)}{\large ). We obtained
the true asymptotic limit of the growth law by performing our simulation for
sufficiently long time. The presence of topologically stable 1/2-disclination
points is reflected in the observed large-momentum dependence k\( ^{-4} \)
of the structure factor. The correlation function was also found to tally with
the theoretical prediction of the correlation function for the two-dimensional
O(2) system. }{\large \par}
\end{abstract}
~~~~~~~~~~~~~~~~~~~~~~~~~~~~~~~~~~~~~~~~~~~~~~~~~~~~~~~~~~~~~~~~~~~~

{\large The phase ordering of various systems with scalar, vector and tensor
order parameters has gained considerable interest over last few years \cite{brayrev}.
The system quenched from a high temperature homogeneous disordered phase into
an ordered phase does not get ordered instantaneously, instead the various degenerate
ground states compete to be selected \cite{brayrev,furrev}. In the process,
the system develops length scales that grow with time and topological defects,
if present, are eliminated. An infinite system will never achieve complete ordering
and the length scale will increase without any bound. If a single growing length
scale characterizes the evolving system then it is said to scale dynamically.
Bray and Rutenberg \cite{energy scaling} proposed a very general technique,
known as the energy scaling approach, to estimate growth laws in purely dissipating
systems that scale dynamically. However their scheme could also be applied to
find out the relation between various length scales for a system in which dynamical
scaling does not hold. If the growth law observed is different from their estimation
then we can say that the system violates dynamical scaling. There are a large
number of systems where the failure of dynamical scaling is observed, e.g. one-dimensional
XY model \cite{1dxy}, the nonconserved two-dimensional O(3) model \cite{2do3},
the conserved spherical model \cite{conigleo} etc. }{\large \par}

{\large The model we have studied is described by the Hamiltonian,}{\large \par}

{\large \[
H=-\sum _{<i,j>}(\phi _{i},\phi _{j})^{2}\]
}{\large \par}

{\large where \( \phi  \) is the usual O(n) vector spins. Due to the spin inversion
symmetry, the model represents a uniaxial nematic. The phase ordering of the
same model was studied by Blundell and Bray \cite{blundell} using a Cell Dynamic
Scheme for d=2, n=2 and d=3, n=3 and in the present work we have studied it
for d=2 and n=3. It differs from the usual two-dimensional O(3) model due to
its local inversion symmetry. While in the two-dimensional O(3) model there
is no stable topological singular defect, in the present model, due to presence
of local inversion symmetry, the order parameter space, instead of being a simple
three dimensional sphere (as in case of usual O(3) models), is a three dimensional
sphere with antipodal points identified. This gives rise to topologically stable
singular point defects of strength \( \pm  \)1/2 (where the director rotates
around the defect core by 180\( ^{o} \)). The mapping of other 1/2 integrals
defects are homotopically equivalent to the mapping of 1/2-defects. A \( - \)1/2
defect configuration is continuously deformable to a 1/2 defect configuration.
Integral defects are topologically unstable due to the so called 'escape to
the third dimension'. Hence the first or the fundamental homotopy group of the
system is just the two element group Z\( _{2} \) (\{0,1\}) \cite{mermin}.}{\large \par}

{\large In general, the O(n) model with d=n-1 supports nonsingular topologically
stable extended spin configurations carrying integral topological charge, known
as topological textures (or anti-textures, for negative topological charge )\cite{thoules}.
In the two-dimensional O(3) model the textures are known as skyrmions, instantons
or baby skyrmions. The various length scales associated with these weakly interacting
textures evolve with different growth laws in the one-dimensional XY model and
in the two-dimensional O(3) model, which give rise to scaling violation in these
systems \cite{1dxy, 2do3}. In the two-dimensional O(3) model the minimum energy
configuration for an isolated texture is obtained by stereographically projecting
the order parameter sphere on the physical space \cite{energy scaling}. The
configuration covers the order parameter space exactly once and hence the texture
is associated with a topological charge 1. In the present model one would expect
the presence of the two-dimensional O(3) like textures, but our effort to find
out them using the algorithm prescribed by Berg and Luscher \cite{berg}resulted
in detecting no textures at all. This may be explained purely on the basis of
homology of the order parameter space. Hindmarsh\cite{hind} on the basis of
topology or more specifically homology of the order parameter space (which is
the projective plane RP\( ^{2} \) in the nematics) have shown that in three
dimensional quenched nematics the probability of occurrence of monopoles is
very low. Unlike in the Heisenberg model, in real nematics in order to get a
monopole the order parameter space has to be covered twice and a special arrangement
over many uncorrelated domains is required. This is responsible for a very low
probability (\( \sim  \)10\( ^{-8} \)) of occurrence of the monopoles. In
case of the two-dimensional RP\( ^{2} \) model (an example of which is the
present model) similar argument should also be valid for the textures and this
perhaps explains why we could not find the textures in this model. In case of
the two-dimensional O(3) model the different growth rates associated with internal
and external length scales of the extended textures are responsible for the
failure of single length scaling \cite{2do3, energy scaling}. Since in the
present model textures (or antitextures) are highly suppressed due to topological
reason, the scaling violation is less likely. In the present paper we have established
that the system scales dynamically. }{\large \par}

{\large We have used Cell Dynamic Scheme\cite{blundell,puri} for studying coarsening
dynamics of the soft spin version of the concerned model. The discrete time
updating relation is, }{\large \par}

{\large \[
\phi _{n+1}(i)=D\left[ \frac{1}{4}\sum _{j}(\widehat{\phi _{n}}(i),\, \widehat{\phi _{n}}(j))\, \phi _{n}(j)-\phi _{n}(i)\right] +E\, \widehat{\phi _{n}}(i)\, tanh(|\phi _{n}(i)|)\]
 The sum is over nearest neighbors of i. The parameter D is called the diffusion
constant, which determines the strength of the coupling between various cells
evolving with time. The value of the parameter E should always be greater that
unity and it determines the depth of quench\cite{puri}. In the above discrete
time updating relation the unit vectors (represented by hats) are used for stability
of the iteration process. However one must avoid using both \( \phi (j) \)
as unit vectors as this leads to a freezing of the configuration in some metastable
region\cite{blundell}.}{\large \par}

\begin{figure}
{\par\centering \resizebox*{0.7\textwidth}{!}{\includegraphics{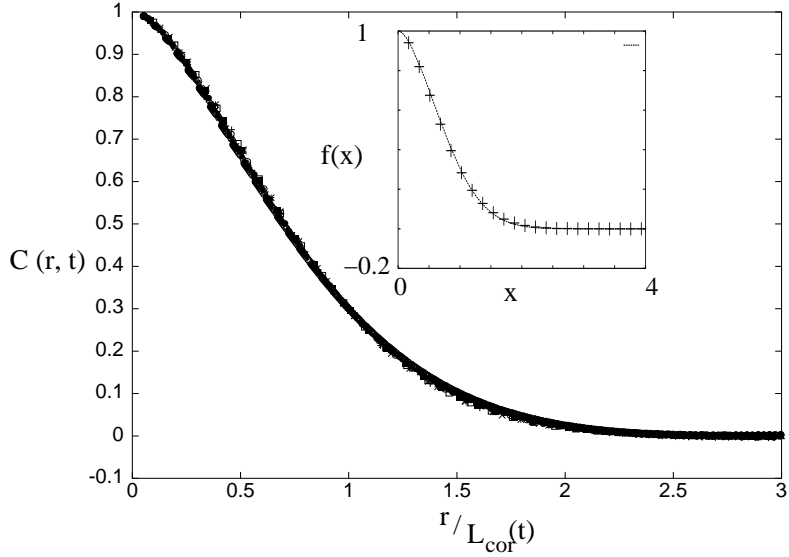}} \par}

\caption{{\large Scaling plot of the correlation function C(r,t) against r/L\protect\( _{cor}\protect \)(t)
for a 256x256 lattice (D=0.1, E=1.1) obtained after collapsing the correlation
function at different time steps (as indicated in figure3). The correlation
length is obtained by using C (L\protect\( _{cor}\protect \)(t),t) = 0.3. The
agreement of Bray Puri prediction \cite{braypuri} for O(2) (\protect\( \cdots \protect \))
model with the scaled correlation function(+) (t=400) is shown in the inset.
The BP function f\protect\( _{BP}\protect \)(x) =(\protect\( 1/\pi )\, \, \protect \)exp(-x\protect\( ^{2}\protect \)/2)~{[}B(1/2,
3/2) {]}\protect\( ^{2}\protect \) F(1/2, 1/2, 2; exp(-x\protect\( ^{2}\protect \))).
The maximum value of the L\protect\( _{cor}\protect \) obtained is 19.35 which
is much smaller the linear size of the lattice, i.e. 256.}\large }
\end{figure}

\begin{figure}
{\par\centering \resizebox*{0.7\textwidth}{!}{\includegraphics{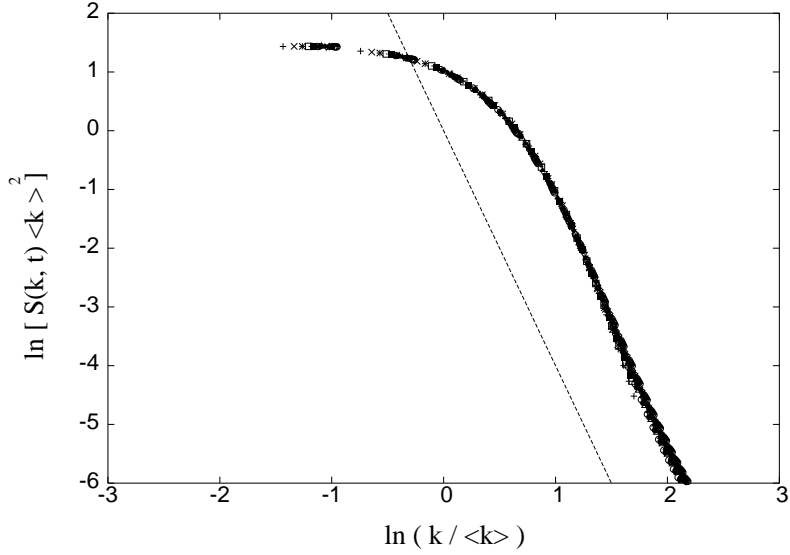}} \par}

\caption{{\large Log-log plot of the structure factor scaling function of a 256x256
lattice (D=0.1, E=1.1). The first moment <k> is used for rescaling of momenta.
The straight line (dotted) has a slope of -4, which indicates the validity of
the generalized Porod's Law. }\large }
\end{figure}

\begin{figure}
{\par\centering \resizebox*{0.7\textwidth}{!}{\includegraphics{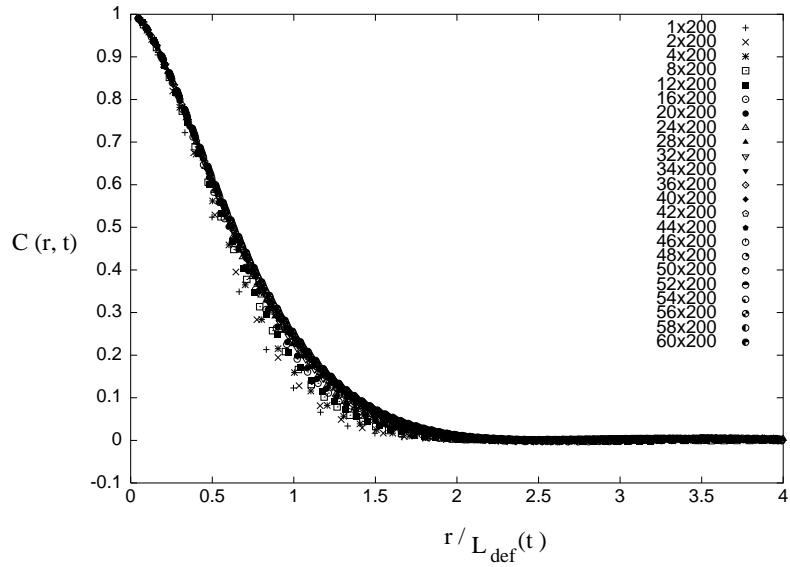}} \par}

\caption{{\large The attempted collapse of the correlation function for a 256x256 lattice
(D=0.1 and E=1.1) with respect to the the defect separation length L\protect\( _{def}\protect \)
\protect\( =\protect \) 1/\protect\( \rho ^{1/2}_{def}\protect \). The failure
of collapse at the initial stages of phase ordering does not indicate the violation
of dynamical scaling. In the asymptotic limit the correlation function scales
well with respect to the defect separation, indicating the proportionality of
L\protect\( _{cor}\protect \)(t) and L\protect\( _{def}\protect \) (t).}\large }
\end{figure}

\begin{figure}
{\par\centering \subfigure{\resizebox*{0.7\textwidth}{!}{\includegraphics{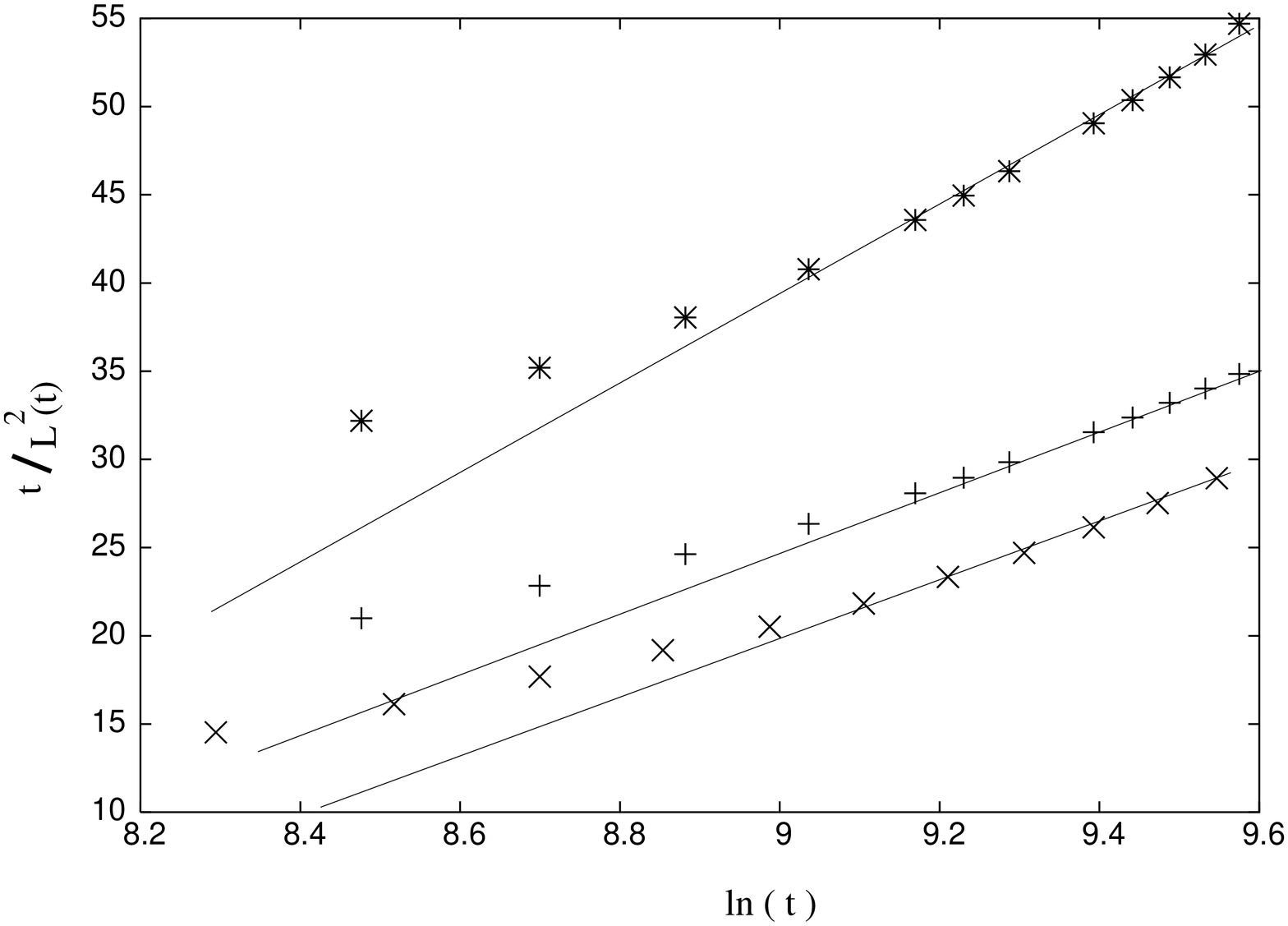}}} \par}

\caption{{\large The plot of t/L\protect\( ^{2}\protect \)(t) vs. ln(t) for three lengths
\protect\( L_{cor}\protect \)(\protect\( \times \protect \)), \protect\( L_{def}\protect \)
(\protect\( +\protect \)) and \protect\( L_{k}\protect \) (\protect\( \star \protect \))
for a 512x512 lattice (D=0.1 and E=1.1). The observed linear dependence at late
times (over a wide range from t=5500 to t=14400) indicates that the dynamical
scaling growth law L(t) \protect\( \sim \protect \) (t/\protect\( ln\protect \)(t/t\protect\( _{0}\protect \)))\protect\( ^{1/2}\protect \)
holds. However the time scale t\protect\( _{0}\protect \) is found to be non-universal. }\large }
\end{figure}

{\large The phase ordering kinetics of the two-dimensional uniaxial nematic
have been studied in details by Zapotocky et al. \cite{ZAPO}. Using a Cell
Dynamic Scheme, they have shown that dynamical scaling is violated in two-dimensional
uniaxial nematics films. They observed different values of the growth exponents,
in the familiar algebraic growth law \( L(t) \) \( \sim  \) \( t^{\phi } \)
(\( \phi  \) is known as growth exponent), corresponding to different length
scales. In determining the effective growth exponents they used the time range
between 200 and 2000. However, as indicated by Rojas and Rutenberg \cite{rojas},
in context of the issue of dynamical scaling in two-dimensional XY model quenched
from above \( T_{KT} \) (the Kosterlitz - Thouless transition temperature),
that in order to decide whether a system violates dynamical scaling or not,
one must find the effective growth exponent in the true asymptotic limit after
it is constant with time and before the finite size effect starts playing its
role. They observed no violation in dynamical scaling in the two-dimensional
XY model. Like integral singular point defects (known as vortices) present in
the two-dimensional XY model, the present two-dimensional model also supports
topologically stable \( \pm 1/2 \) disclination points. So it is expected that
the growth laws should be similar and this is the main finding of the present
work. In figure 5, we have shown the Schlieren patterns in a 180x180 uniaxial
nematic placed between crossed polarizers at different times indicated in the
figure. The patterns were obtained in the same way as discussed in references
\cite{ZAPO, MULLER, BERGE}. By performing the simulation for sufficiently long
run to get the true asymptotic limit, in the present work we have shown that,
the two-dimensional uniaxial nematic scales dynamically by establishing that
the same asymptotic growth law is valid for various length scales. Instead of
the usual \( t^{\phi } \) growth law, the system was found to scale asymptotically
in a manner similar to the two-dimensional XY model quenched from above \( T_{KT} \)
with the growth law, L(t) \( \sim  \)} \textbf{\large \( \, \, (t/ln(t/t_{0}))^{1/2} \)}
{\large ( \( t_{0} \) nonuniversal time scale ) \cite{rojas, tkt}. We have
performed our simulation with two lattices sizes 256x256 and 512x512. By comparing
the results of the two lattice sizes, we did not find any significant finite
size effect upto the time limit we have investigated. }{\large \par}

{\large The normalized correlation function in the present model is given by,}{\large \par}

{\large \[
C(r,t)=3/2<(\widehat{\phi }(0),\widehat{\phi }(r))^{2}>-1/2\]
where < > represents the average over various random initial states (random
length and magnitude). The scaling form of correlation function is given by, }{\large \par}

{\large \[
C(r,t)=f(r/L_{cor}(t))\]
Where the \( L_{cor}(t) \) is the length scale required to collapse the correlation
functions for different time. In the Figure1 we have shown the scaling plot
of C(r,t) averaged over 20 initial states for a 256x256 lattice. }{\large \par}

{\large The structure factor scales with respect to \( L_{k}=1/<k> \)\cite{rojas},
where <k>} = {\large \( \sum S(k,t)\, K/ \) \( \sum S(k,t) \) , is the first
moment of structure factor. The scaling form of the the structure factor is
given by,}{\large \par}

{\large \[
S(k,t)=L_{k}^{2}\, g(kL_{k}(t))\]
In Figure2 we have shown the plot of ln(L\( _{k} \)\( ^{-2} \)S(k,t)) against
(kL\( _{k} \)). From generalized Porod's law the large-momentum structure factor
for a two-dimensional system with point defects should be proportional to \( \rho _{def}\, k^{-4} \)
\cite{brayrev,ZAPO, porod, braypuri}. In the present system the density of
the point defects \( (\rho _{def}) \) scales as L\( _{def} \)\( ^{-2} \),
where L\( _{def} \) is the typical defect separation length. In the large-momentum
limit we obtained the the slope of ln(L\( _{k} \)\( ^{-2} \)S(k,t)) versus
kL\( _{k} \) plot equal to \( - \)4 as shown in the Figure2, which verifies
Porod's Law. The good collapse of the tail, verifies that L\( _{k} \) and L\( _{def} \)
have asymptotically the same growth law. In Figure3, where we have tried the
collapse of correlation function with respect to the defect separation, poor
collapse at the initial stage of dynamics however does not indicate the violation
of dynamical scaling. At late stages the collapse shown in Figure3 is quite
acceptable. In Figure4 we have verified that the growth law for all concerned
lengths are asymptotically same as that for the two-dimensional XY model, because
of the linear dependence of t/L\( ^{2} \)(t) on ln(t) . Figure4 also verifies
that the present system does not violate dynamical scaling. The correlation
function that we obtained in our simulation also tallies with the Bray Puri
prediction \cite{braypuri} of the equal time correlation function for O(2)
model, i.e. }{\large \par}

{\large \[
C(r,t)=(\gamma /\pi )[B(1/2\, ,\, 3/2)]^{2}F(1/2\, ,\, 1/2\, ,\, 2\, ;\, \gamma ^{2})\]
}{\large \par}

{\Large where,} \textbf{\large \[
\gamma (r,t)=exp(-r^{2}/8t)\]
}{\large \par}

{\large and B(x,y) is the beta function, and F(a, b, c; z) is the hypergeometric
function. }{\large \par}

{\large Thus in the scaling form we have C(r,t)=f\( _{BP} \)(x) (BP stands
for Bray and Puri), where x=r/L(t) and L(t)=(4t)\( ^{1/2} \). The logarithmic
factor is not correctly reflected in that function. However for comparison we
have plotted the f\( _{BP} \)(x) and the scaled correlation function (for t=400)
in the inset of Figure1. We have performed our simulation with higher values
of D(0.5) and got similar asymptotic results. Higher values of D are useful
in achieving the asymptotic regime faster. However the finite size effect is
also more prominent in case of large D. We have not considered noise in the
time evolution equation, hence are effectively working at T=0. However it is
known that quenching to T=0 may lead to metastable freezing \cite{yurke}. In
order to check that our results are not influenced by such freezing, we performed
a number of simulations (almost 100 steps) with noise, by adding a constant
amplitude (of the order of 0.1) random configuration to the order parameter.
The noise amplitude used was enough for generating large number of pairs of
disclination point defects. We could not find any discrepancy with the results
obtained without noise.}{\large \par}

{\large To summarize we would like to focus on the main findings of the paper.
We have confirmed that dynamical scaling is not violated in two-dimensional
nematic with order parameter dimensionality three and asymptotically the growth
laws are same as that of the two-dimensional XY model quenched from above T\( _{KT} \)
(i.e. the initial state with free vortices). Consideration of topological defect
in the issue of dynamics is very necessary, because the structure of the defects
determines the large-momentum dependence of the structure factor, which has
an important role in the determination of the growth laws \cite{energy scaling}.
In the O(n) model with n\( \leq  \)2, the topological defects dominate the
dynamics. Since both the present two-dimensional model and the two-dimensional
XY model support singular point defects, it is expected that the dynamics should
be similar and this is established in the present work. In the present work
we were able to establish the expected result and achieved the true asymptotic
limit of the dynamical scaling by performing longer simulation with the help
of increased computational power now available. }{\large \par}

\textbf{\large Acknowledgments:} {\large The authors are thankful} \textbf{\large }{\large to
Dr. A.D. Rutenberg for useful discussions. One of us (S. Dutta) acknowledges
the financial assistance from the} \textbf{\large COUNCIL OF SCIENTIFIC AND
INDUSTRIAL RESEARCH} {\large (}\textbf{\large CSIR}{\large ), India. }{\large \par}

\end{document}